\documentclass[12pt,a4paper]{article}

  \usepackage{a4wide}
  \usepackage{latexsym}
  \usepackage{epsf}
  \usepackage{amssymb}
  \usepackage{graphicx}
  \usepackage{amsmath, cite}
  \usepackage{amsmath,amssymb,amsthm}
  \usepackage{verbatim}
  \usepackage{hyperref}

\renewcommand{\d}{\textrm{d}}
\newcommand{\e}{\textrm{e}}

\newcommand{\w}{\wedge}

\newcommand{\be}{\begin{equation}}
\newcommand{\ee}{\end{equation}}
\newcommand{\ba}{\begin{eqnarray}}
\newcommand{\ea}{\end{eqnarray}}

\renewcommand{\d}{\textrm{d}}

\begin{document}
\numberwithin{equation}{section}

\begin{center}

\begin{flushright}
{\small UUITP-07/14
 \\ 
} \normalsize
\end{flushright}
\vspace{0.9 cm}

{\LARGE \bf{Localised anti-branes in non-compact\vspace{0.5cm}}}

{\LARGE \bf{throats at zero and finite $T$\vspace{0.2cm}}}

\vspace{1.1 cm} {\large J. Bl{\aa }b\"ack$^a$,  U.H. Danielsson$^a$, D. Junghans$^b$, \\\vspace{0.2cm}
T. Van Riet$^c$, S.C. Vargas$^a$}\\

\vspace{0.8 cm}{$^{a}$ Institutionen f{\"o}r fysik och astronomi,\\ Uppsala Universitet, Uppsala, Sweden}\\
\vspace{.1cm}{$^{b}$ Center for Fundamental Physics $\&$ Institute for Advanced Study,\\
Hong Kong University of Science and Technology, Hong Kong}\\
\vspace{.1 cm} {$^c$ Instituut voor Theoretische Fysica, K.U. Leuven,\\
Celestijnenlaan 200D B-3001 Leuven, Belgium} \footnote{{\ttfamily {johan.blaback,  ulf.danielsson, sergio.vargas @ physics.uu.se, daniel@ust.hk, thomasvr @ itf.fys.kuleuven.be}}}\\

\vspace{1.4cm}

{\bf Abstract}
\end{center}

\begin{quotation}
We investigate the 3-form singularities that are typical to anti-brane solutions in supergravity and check whether they can be cloaked by a finite temperature horizon. For anti-D3-branes in the Klebanov-Strassler background, this was already shown numerically to be impossible when the branes are partially smeared. In this paper, we present analytic arguments that also localised branes remain with singular 3-form fluxes at both zero and finite temperature. These results may have important, possibly fatal, consequences for constructions of meta-stable de Sitter vacua through uplifting.
\end{quotation}
\newpage

\section{Introduction}

The explicit breaking of supersymmetry (SUSY) by brane sources in warped throats could be a natural mechanism to create meta-stable vacua in string theory with a tunable amount of SUSY-breaking and hence perturbative control. This can be used for the construction of de Sitter vacua \cite{Kachru:2003aw}, brane inflation \cite{Kachru:2003sx}, holographic studies of SUSY-breaking in gauge theories \cite{Kachru:2002gs, Argurio:2007qk, Klebanov:2010qs} or the construction of non-extremal black hole micro-states \cite{Bena:2011fc}.

The prime example for this mechanism is to insert an anti-D3-brane at the tip of the Klebanov-Strassler (KS) geometry \cite{Klebanov:2000hb, Kachru:2002gs}. As is well known by now, the first attempts to construct the corresponding supergravity solution \cite{McGuirk:2009xx, Bena:2009xk} revealed that its infra-red region has a diverging 3-form flux density\footnote{This is the result in Einstein frame, but also the string frame flux density is divergent.} 
\begin{equation}
\e^{-\phi} |H_3|^2 \rightarrow \infty,
\end{equation} 
where $H_3=dB_2$ is the field strength of the NSNS $B$-field.  At the time of \cite{McGuirk:2009xx, Bena:2009xk}, this result was preliminary (although the singularity could have been anticipated from \cite{Polchinski:2000uf}) since the supergravity solutions were constructed using various approximations, i.e., a partial smearing of the anti-branes and a linearisation of their perturbation of the BPS background. In the last years, however, it became clear that flux singularities are both ubiquitous and on a firm footing within the supergravity approximation. For anti-D3-branes in the KS background, they were demonstrated to neither be an artifact of the linearisation \cite{Massai:2012jn, Bena:2012bk} nor of the partial smearing \cite{Gautason:2013zw}. Furthermore, several studies of other setups with anti-branes in flux backgrounds led to similar results \cite{Giecold:2011gw, Bena:2010gs, Massai:2011vi, Blaback:2011nz, Blaback:2011pn, Giecold:2013pza, Cottrell:2013asa, Blaback:2013hqa, Bena:2014bxa}. A review of various anti-brane solutions and their physics with a complete list of references will appear soon \cite{toappear1}.

Although the construction of explicit supergravity solutions with SUSY broken by anti-branes is very involved, it has been realised that this is not required to find out whether the 3-form singularities will occur. Instead, it is often possible to formulate so-called `no-go theorems' for the existence of solutions with regular fluxes. The first such no-go theorem was found for anti-D6-branes in massive type IIA \cite{Blaback:2011nz}. This was later generalised to anti-D$p$-branes smeared over $6-p$ directions \cite{Bena:2013hr} including finite temperature $T$ (see also \cite{Bena:2012bk}), to smeared anti-M2-branes at finite $T$ \cite{Blaback:2013hqa} and to localised anti-D3-branes in compact geometries with a KS-like throat at $T=0$ \cite{Gautason:2013zw}.

It should not be any surprising that such no-go theorems can be found without knowing the solutions since there exists a very straightforward intuition that explains the nature of the singular fluxes \cite{DeWolfe:2004qx, Blaback:2010sj, Blaback:2011nz, Blaback:2012nf}: since the fluxes carry charges opposite to that of the anti-brane, they are gravitationally and electromagnetically attracted towards the anti-brane resulting in an increased pile-up of flux. The only force that can counterbalance this (otherwise fatal \cite{Blaback:2012nf, Kachru:2002gs}) attraction is due to the gradient energy in the flux. If it is strong enough to generate a balance of forces, one would expect an increased but finite flux density. This is not the case in all examples studied so far where SUSY is broken by the anti-brane.

In the case of anti-D3-branes, the energy density is integrable \cite{Bena:2009xk} and does therefore not immediately invalidate anti-branes as an uplifting mechanism \cite{Junghans:2014xfa}. Furthermore, the solution is well-behaved in the UV and stands some very non-trivial tests \cite{Dymarsky:2011pm, Dymarsky:2013tna}. Nevertheless, in order for the solution to be physical, the singularities arising in the supergravity approximation should be resolved by some mechanism in string theory. A natural proposal for such a mechanism is brane polarisation via the Myers effect \cite{Myers:1999ps}. If an anti-D$p$-brane polarises into a spherical D$(p+2)$-brane with worldvolume flux carrying the anti-D$p$ charge, then one expects a finite pile-up of the bulk fluxes since their attraction to the brane is softened as the anti-brane charge is spread out over a sphere instead of a point \cite{Polchinski:2000uf}. Indeed, brane polarisation has been shown to resolve flux singularities in both SUSY and non-SUSY anti-D6-brane solutions with an AdS worldvolume \cite{Junghans:2014wda} (see also \cite{Apruzzi:2013yva, Gaiotto:2014lca}). However, this does not seem to happen in setups in which the curvature of the anti-brane worldvolume is either zero (as in the non-compact holographic backgrounds of \cite{Kachru:2002gs, Klebanov:2010qs}) or hierarchically small and positive (as in the string cosmology context of \cite{Kachru:2003aw}). In all such setups studied so far, brane polarisation can explicitly be shown not to occur \cite{Bena:2012tx, Bena:2012vz, Bena:2014bxa}, although some polarisation channels are not fully understood yet. 
In \cite{Blaback:2012nf}, it was argued that the flux singularities might then rather indicate that the anti-brane states are not meta-stable but decay perturbatively to the SUSY vacuum through brane-flux annihilation.

Alternatively, one might wonder whether other stringy effects could lead to well-behaved anti-brane solutions beyond the supergravity regime. In the absence of a full string theory solution, however, it appears hopeless to identify the correct stringy mechanism that resolves the singularities or, conversely, show explicitly that all possible mechanisms fail. A more promising idea is to try to find a general argument for why or why not the singularities are expected to be resolved in string theory. Such a general criterion for distinguishing `good' from `bad' singularities was formulated by Gubser \cite{Gubser:2000nd}. It states that good singularities (i.e., those that are a mere artifact of using classical supergravity) can be cloaked behind a horizon if sufficient temperature is turned on.
If reliable, this criterion is very powerful as it allows to bypass the difficult question which stringy effect could be responsible for the resolution of a singularity. Instead, one can decide whether a singularity is benign by simply looking for a finite temperature resolution of the singular solution. However, for localised anti-D3-branes in the Klebanov-Tseytlin (KT) throat and  partially smeared anti-D3 branes in the Klebanov-Strassler (KS) throat, it was shown numerically by Bena et al. \cite{Bena:2012ek} (see also \cite{Buchel:2013dla, Bena:2013hr}) that such a resolution does not occur.  This is a strong indication that the IR singularity is genuinely `bad'.

Nevertheless, one could worry that 1) numerics are not sufficient as a proof, 2) for localised anti-branes in KS the result can differ, or 3) for throats different than KS it might work. In this paper, we therefore extend the above mentioned no-go theorems to include these concerns. In particular, we present analytic arguments that, for fully localised anti-branes, turning on a finite temperature does not resolve the singularities. Our arguments apply to anti-D3-branes in the KT and KS geometries as well as to analogous setups in other dimensions. We first discuss the case of anti-D3-branes in KT for which we extend a simple argument presented in \cite{Blaback:2011nz, Bena:2013hr}. This yields a no-go theorem excluding regular anti-brane solutions in this background at both zero and finite temperature. In order to treat the more complicated case of anti-D3-branes in KS, we then generalise the existing no-go theorem of \cite{Gautason:2013zw} in two ways. First, we derive a version of the no-go that is valid for non-compact throat geometries. Our key observation is that the equations of motion then relate the IR boundary conditions for the fields at the anti-D3 position to a boundary term in the UV, which can be shown to equal the ADM mass at $T=0$. This relation implies that, whenever the number of anti-D3-branes (and, hence, the ADM mass) is non-zero, a singular flux is generated in the IR. Second, we argue that this result can be extended to finite temperature. Under certain conditions, the boundary term is then still related to the ADM mass, which can be shown to lead to a singular flux at the horizon. This may indicate that the Gubser criterion is violated for anti-D3-branes in KS.

The rest of this paper is organised as follows. In section \ref{antikt}, we show analytically that the flux singularity is not cloaked by finite $T$ in the case of anti-D3-branes in KT. In sections \ref{review} and \ref{generalised-nogo}, we discuss a formulation of the no-go theorem that extends to KS, both at zero and at finite $T$. Finally, in section \ref{discussion}, we discuss our results.

\section{Anti-D3-branes in Klebanov-Tseytlin at finite $T$ }
\label{antikt}

In reference \cite{Blaback:2011nz}, a simple argument was given to prove the 3-form singularity for anti-D6-branes inserted in a flux background for which the charge dissolved in the flux has the opposite sign of the anti-D6-brane charge far away from the latter. This argument was then extended in \cite{Bena:2013hr} to general anti-D$p$-branes in flux backgrounds with opposite charge including finite temperature, where the space transversal to the branes was taken to be $\mathbb{T}^{6-p}\times \mathbb{R}^3$ and the branes were smeared over $\mathbb{T}^{6-p}$.
As a warm-up, we now demonstrate that this technique can readily be applied to localised anti-D3-branes in the KT background \cite{Klebanov:2000nc}. For the latter case, it was numerically shown in \cite{Bena:2012ek} that no regular solution exists, and our analytic proof below confirms this result. The key to our argument is that the symmetries of the setup allow to reduce the equations of motion to ordinary differential equations that only depend on one radial variable $r$. This then leads to a strong constraint on the behaviour of one of the fields, which cannot be satisfied in the presence of anti-D3-branes at the tip unless a singularity forms at the horizon. Since the KS throat has less symmetries, it is not possible to extend this technique to that background, and so we will use the more powerful and slightly more abstract formalism of \cite{Gautason:2013zw} in the next section.

The Ansatz for the metric in Einstein frame describing anti-D3-branes in KT is  \cite{Bena:2012ek}
\begin{equation}
\d s^2_{10} = \e^{2\tilde{A}(r)} g_{\mu\nu} \d x^\mu \d x^\nu + \e^{2B(r)-2f(r)} \d r^2 + \e^{2 B(r)}  g_5^2 + \e^{2 {C}(r)}\left(g_1^2 + g_2^2 + g_3^2 + g_4^2 \right)\,,
\end{equation}
where $g_{tt} =-\e^{2f(r)}$ and $g_{ij}=\delta_{ij}$ for $i,j=1,2,3$.\footnote{Here, we have put a tilde on the warp factor $\tilde A$ in order to distinguish it from a different function $A$ that appears in the Ansatz of \cite{Bena:2009xk} and in appendix \ref{ssec:appa1}. From section \ref{review} on, we will denote the warp factor by $A$ as usual.} $f(r)$ is a blackening factor, which becomes non-zero if a finite temperature is turned on and diverges at the horizon at finite $r$. The one-forms $g_1,\ldots, g_5$ are the one-forms of $T^{1,1}$ used, for example, in \cite{Bena:2009xk}, and some basic identities for these forms appear in appendix \ref{AppConventions}. The flux Ansatz is
\begin{equation}
\begin{split}
F_5 &= -(1+\star_{10}) \e^{-4\tilde{A}(r)-f(r)}\star_6 \d \alpha(r)\,,\\
H_3 &= \lambda(r)\,\e^{\phi(r)}\star_6 F_3\,,\\
F_3 &= P\, (g_1 \w g_2 + g_3 \w g_4) \w g_5\,,
\end{split}
\end{equation}
where $P$ is a constant.
This Ansatz is a special case of the KS Ansatz used in appendix \ref{ssec:appa1} with a simple relabelling of the fields.

From the $H_3$ equation of motion
\be
\d(\star_{10}\e^{-\phi}H_3) = -F_5\wedge F_3\,,
\ee
we deduce
\be
\lambda \e^{4\tilde{A}+f} = \alpha +\alpha_0\,,
\ee
with $\alpha_0$ an integration constant that can be set to zero using a $C_4$ gauge transformation.  Together with the $F_5$ Bianchi identity away from the source position
\be
\d F_5 = H_3\wedge F_3\,,
\ee
we find the second-order differential equation
\be
\left( \e^{-4\tilde{A}+4{C}} \alpha'\right)' = 2P^2 \e^{\phi -4\tilde{A}-2f} \alpha\,,
\ee
where a prime denotes a derivative w.r.t.~$r$. One then observes that this equation implies
\be\label{derivativesigns}
\text{sgn}(\alpha) = \text{sgn}(\alpha'')\qquad\text{when}\qquad \alpha'=0.
\ee
This condition is sufficient to demonstrate that a regular solution cannot exist, following the reasoning in \cite{Blaback:2011nz, Bena:2013hr}. In fact, it implies that the 3-form singularity is not cloaked by the horizon but instead replaced to live at the position of the latter. For convenience, we briefly summarise the arguments of \cite{Blaback:2011nz, Bena:2013hr}. One can demonstrate that the sign of the anti-D3 charge is determined by the sign of  $\alpha'$ at the horizon,
\be\label{charge}
\text{sgn}{Q} = \text{sgn}(\alpha')|_{\text{horizon}}\,,
\ee
where we take conventions in which anti-branes have negative charge.
On the other hand, if we demand that the anti-brane lives in a flux background that has positive charge at large distance (i.e., the fluxes become ISD at infinity), we have
\be \label{signflux}
\alpha > 0 \qquad \text{at large radius}\,.
\ee
It is then easy to demonstrate that
\be \label{nonzero}
\alpha(\text{horizon})\neq 0\,.
\ee
If $\alpha$ were instead zero at the horizon, the conditions (\ref{derivativesigns}) and (\ref{charge}) would imply a discrete jump in $\alpha$ and, hence, a singularity between the horizon and infinity. However, (\ref{nonzero}) implies that the 3-form flux density
\be\label{KTfiniteT}
\e^{-\phi}|H_3|^2 = 2P^2 \alpha^2 \e^{\phi-8 \tilde A-2f-2{B}-4{C}}
\ee
is singular at the horizon since the blackening factor $\e^{f}$ approaches zero there, whereas all other quantities remain finite. Note that this is not the case if one puts a D3-brane instead of an anti-D3-brane at the tip since the conditions \eqref{derivativesigns}, \eqref{charge} and \eqref{signflux} are then compatible with $\alpha=0$ at the horizon. It is also well-known that the standard metric singularity is then cloaked by the finite temperature horizon.

\section{A review of the existing no-go}
\label{review}

In this section, we review the no-go technique of \cite{Gautason:2013zw}, which applies to anti-D3-branes in compact geometries with a KS throat and to a number of other setups for which the fields satisfy a similar Ansatz. In the next section, we will extend the argument to non-compact geometries and discuss its generalisation to finite temperature. Since the results of \cite{Gautason:2013zw} are more general than those of \cite{Blaback:2011nz, Bena:2013hr}, the formalism is more abstract. However, when applied to known cases such as anti-D3-branes in KT or the anti-D6-brane models of \cite{Blaback:2011nz, Blaback:2011pn, Bena:2013hr}, both techniques yield exactly the same equations.

In the following, we will consider flux compactifications of type II supergravity to $p+1$ dimensions with spacetime-filling D$p$-branes and O$p$-planes. A general Ansatz for the metric is then
\begin{equation}
\d s^2_{10} = \e^{2A} \tilde g_{\mu\nu} \d x^\mu \d x^\nu + \d s^2_{9-p}\,,
\end{equation}
where $\mu,\nu =0,\ldots,p$ and $A$ is a function of the internal coordinates.
The essential ingredient of the no-go theorem of \cite{Gautason:2013zw} is the following relation (inspired by the work of \cite{Burgess:2011rv}) between the cosmological constant $\Lambda$ in $p+1$ dimensions, the on-shell actions of the D-branes and O-planes and a term due to topological fluxes:
\begin{equation}\label{mastereq1}
\frac{8 v \mathcal{V}}{p-1} \Lambda = \left(1+\frac{p-3}{2}c\right) \left[S^{(p)}_\textrm{DBI} + S_\textrm{WZ}^{(p)}\right] + \int \mathcal{F}(c).
\end{equation}
As detailed in \cite{Gautason:2013zw}, this relation follows directly from the equations of motion and is a consequence of scaling symmetries of the supergravity action.

The notation used in \eqref{mastereq1} is as follows. The number $c$ is a ``gauge'' which can be chosen freely, and $v$ and $\mathcal{V}$ are the volume factors 
\begin{equation}
v = \int \tilde \star_{p+1} 1 \, , \qquad \mathcal{V} = \int \star_{9-p}\, \e^{\left( p -1 \right) A} \,,
\end{equation}
where $v$ cancels out in (\ref{mastereq1}) with implicit volume factors on the right-hand side. Here and in the following, we denote the Hodge star constructed from $\tilde g_{\mu\nu}$ by $\tilde \star_{p+1}$, the one constructed from the warped metric $\e^{2A} \tilde g_{\mu\nu}$ by $\star_{p+1}$ and the one constructed from the internal part of the metric by $\star_{9-p}$. The form $ \mathcal{F}(c)$ is a certain combination of bulk fluxes, which we will make explicit below for a certain choice of $c$. The other terms on the right-hand side of (\ref{mastereq1}) are the on-shell DBI and WZ actions of the D-branes and O-planes,
\begin{align}\label{braneaction}
& S^{(p)}_\textrm{DBI} = \mp\mu_p \int \star_{p+1}\e^{\tfrac{p-3}{4}\phi}\wedge \sigma(\delta_{9-p})\,,\nonumber\\
& S_\textrm{WZ}^{(p)}  = \pm \mu_p\int C_{p+1}\wedge \sigma(\delta_{9-p}),
\end{align}
where the upper sign is for D$p$-branes, the lower one for O$p$-planes and $\mu_p$ is the absolute value of the D$p$/O$p$ charge. The operator $\sigma$ acts on $p$-forms by reverting all indices (see appendix \ref{AppConventions}). We have neglected couplings to other bulk fields and worldvolume fields since they are not relevant in the context of this paper.

Let us now specialise to a compact space with a warped throat region that is locally described by the KS solution as in \cite{Giddings:2001yu}. In order to cancel the tadpole due to the D3-brane charge of the 3-form fluxes, such a solution requires the presence of O-planes in the bulk, where we will restrict to the simple case of a compactification with O3-planes in the following (more generally, one could also have D7-branes and O7-planes in the bulk, which can also carry D3-brane charge). Furthermore, we put an anti-D3-brane at the tip of the KS throat in order to make contact with the KKLT scenario \cite{Kachru:2003aw}.

Following \cite{Gautason:2013zw}, we can then make the following Ansatz for the fields:
\begin{align} \label{ansatz-ks}
\hat C_4&= \tilde \star_4 \alpha\,,\nonumber\\ F_5&=-(1+\star_{10})\e^{-4A}\star_6\d\alpha\,,\nonumber\\
H_3&= \e^{\phi-4A}\star_6(\alpha F_3 + X_3)\,,
\end{align}
where $\hat C_4$ denotes the spacetime-filling part of the RR potential $C_4$.
The dilaton $\phi$ and the gauge potential profile $\alpha$ are functions on the internal space. Furthermore, the 3-form fluxes $H_3$ and $F_3$ are both closed, and $X_3$ is a closed 3-form that is not proportional to $F_3$. One can then show that (\ref{mastereq1})
implies \cite{Gautason:2013zw}
\begin{equation}\label{mastereq2}
\Lambda = -\tfrac{1}{4\mathcal{V}}{\bar N}\mu_3(\e^{4A_0}+\alpha_0) + \tfrac{1}{16 \mathcal{V}}N_{O3}\mu_3(\e^{4A_*}-\alpha_*) - \tfrac{1}{4\mathcal{V}}\int_6  H_3\wedge X_3 +\Lambda_{np}\,,
\end{equation}
where $\e^{4A_0}, \alpha_0$ ($\e^{4A_*}, \alpha_*$) denote the values of the warp factor and the gauge potential profile at the position of the anti-D3-branes (O3-planes) and ${\bar N}, N_{O3}$ are the anti-D3-brane and O3-plane numbers, respectively. By adding the term $\Lambda_{np}$ on the right-hand side, we have also incorporated the effect of non-perturbative corrections to the 4D effective scalar potential, which is necessary to find a de Sitter vacuum in the KKLT scenario \cite{Kachru:2003aw}.

Assuming that the fields approach the BPS background of \cite{Giddings:2001yu} far away from the anti-D3-brane, one can set $\e^{4A_*}=\alpha_*$ since the O3-planes are outside the warped conifold region. Thus, the O3-plane contribution in (\ref{mastereq2}) vanishes. One can show \cite{Gautason:2013zw} that the same assumption implies that $X_3$ is exact\footnote{For the partially smeared solution of \cite{Bena:2009xk}, this is explicitly shown in appendix \ref{Secbandm}.}, and, as a consequence, $\int_6  H_3\wedge X_3 = 0$. Furthermore, the DBI term $\e^{4A_0}$ vanishes since $\e^{4A}\rightarrow 0$ near the anti-brane. Hence, \eqref{mastereq2} reduces to
\begin{equation}\label{lambdacompact}
\Lambda = -\tfrac{1}{4\mathcal{V}}{\bar N}\mu_3\alpha_0 + \Lambda_{np}\, .
\end{equation}
In the absence of an uplifting term due to anti-D3-branes (i.e., for $\alpha_0 = 0$), $\Lambda_{np}$ can be shown to be \emph{negative} \cite{Kachru:2003aw}. A dS solution with $\Lambda > 0$ as in \cite{Kachru:2003aw} therefore requires $\alpha_0$ to be non-zero.\footnote{The term $\Lambda_{np}$ as defined in \cite{Gautason:2013zw} contains an implicit dependence on $\alpha_0$ such that $\alpha_0$ must be \emph{positive} in order to obtain a dS solution (see \cite{Junghans:2014xfa} for a discussion of this subtlety). This is consistent with our explicit result for $\alpha_0$ in section \ref{generalised-nogo}.}

Similar to the KT example treated above, this is sufficient to show the presence of divergent 3-form flux near the source. This can be seen as follows. From our Ansatz \eqref{ansatz-ks}, we deduce the $H_3$ energy density close to the anti-D3-brane,
\begin{equation}\label{inequality}
\e^{-\phi}|H_3|^2 = \e^{\phi-8A}|\alpha F_3 +  X_3|^2 = \e^{\phi-8A}\alpha^2 |F_3|^2 + \ldots = \e^{\phi-2A}\alpha^2 m^2\, + \ldots, 
\end{equation}
where $m$ is the $F_3$ flux quantum at the tip and the dots denote possible additional terms due to $X_3$ and exact terms in $F_3$. Note that $m$ is non-zero in the KS solution in order to prevent the 3-sphere inside of the deformed conifold from shrinking to zero size at the tip \cite{Klebanov:2000hb}. For the last equality, we have assumed that the metric of the space transverse to the anti-D3-brane is locally of the form $g_{mn} = \e^{-2A} \tilde g_{mn}$ near the anti-brane, where $\tilde g_{mn}$ denotes a regular metric that we need not specify further.
Since $\alpha_0$ is finite but $\e^{-2A}$ grows to infinity towards the anti-D3 position, we find that the $H_3$ flux density diverges at least as bad as{\footnote{For a regular dilaton, the dilaton equation then implies that also the $F_3$ energy density is divergent \cite{Gautason:2013zw}. This is in agreement with the explicit results of \cite{McGuirk:2009xx, Bena:2009xk}.}
\begin{equation}\label{sing}
\e^{-\phi}|H_3|^2\sim \e^{-2A}\,.
\end{equation}
Under the above assumptions, this conclusion is independent of whether the near-brane behaviour of $\e^{2A}$ is the same as in the standard D3-brane solution in flat space or whether, as recently proposed in \cite{Bena:2014bxa}, it diverges more strongly in the KS throat.

One might wonder whether other terms in the $H_3$ energy density could somehow combine to cancel the singular term $\propto \e^{-2A}$ on the right-hand side of \eqref{inequality}. Since the internal metric is positive definite, different components of $H_3$ cannot cancel out against each other in the energy density. Hence, in order that no singular term appears in \eqref{inequality}, $X_3$ would have to cancel $\alpha F_3$ at the anti-brane for every component individually. In all explicitly known cases, however, such a cancellation does not happen. In particular, one can verify this in the non-linear solution for partially smeared anti-D3-branes described in \cite{Bena:2012bk, Bena:2012vz}. The component of $X_3$ along the 3-sphere then vanishes at the tip while the one of $F_3$ is non-zero and proportional to $m$. Hence, the terms represented by the dots in \eqref{inequality} are manifestly positive and cannot cancel the singular term.

\section{The generalised no-go theorem}
\label{generalised-nogo}

In what follows, we generalise the master equation (\ref{mastereq1}) found in \cite{Gautason:2013zw}, which is valid if the solution is maximally symmetric in the $(p+1)$-dimensional spacetime parallel to the anti-branes and the transversal space is compact. In order to analyse anti-branes in non-compact geometries at both zero and finite $T$, we will have to drop these two assumptions. Our derivation follows the strategy of the appendix of \cite{Gautason:2013zw}, where (\ref{mastereq1}) was obtained by integrating a certain combination of the equations of motion over the compact space. Unlike \cite{Gautason:2013zw}, however, we now include a blackening factor in the spacetime metric, which becomes non-trivial at finite temperature and breaks maximal symmetry. Furthermore, we keep track of all total derivative terms in the equations of motion, which could be neglected in \cite{Gautason:2013zw} since they integrate to zero on a compact space. In the non-compact geometries we consider, the total derivatives instead integrate to a boundary term, which is non-zero in general and plays a central role in our discussion. As we will explain below, it is expected to equal the ADM mass at zero temperature if the brane worldvolume is flat (i.e., for $\Lambda=0$). Our generalised master equation then relates this to a non-vanishing gauge potential at the anti-brane horizon, which in turn implies a singularity in the $H_3$ energy density. At finite temperature, we find that the boundary term is the sum of a term related to the zero-temperature ADM mass and a term proportional to $TS$, where $S$ denotes the entropy associated to the black hole horizon. Under certain conditions, we then again find a no-go theorem against the existence of solutions with regular flux at the horizon.

\subsection{The master equation}
\label{section-mastereq}

Let us consider type II supergravity with a metric of the form
\begin{equation}\label{metricANSATZ}
\mathrm{d} s^2_{10} = \e^{2A} g_{\mu\nu} \d x^\mu \d x^\nu + \mathrm{d} s^2_{9-p}\,,
\end{equation}
where $g_{tt}$ contains the blackening factor $\e^{2f}$.  The metric with the blackening factor taken off is denoted by $\tilde{g}_{\mu\nu}$ from here on, and we assume that $\tilde{g}_{\mu\nu}$ is a maximally symmetric space with cosmological constant $\Lambda$.\footnote{It is unclear to us whether this finite-$T$ Ansatz is sensible when $\Lambda\neq 0$, but this will not play any important role in this paper, where the finite-$T$ case is studied for $\Lambda=0$.} We furthermore assume that the $p$-brane worldvolume extends along the first $p+1$ directions and that there are no fields with a non-trivial profile along those directions, apart from possible form fields filling all the worldvolume directions.  The non-zero form field strengths in the backgrounds we consider are $H_3, F_{p+2}, F_{6-p}$, where $F_{p+2}$ is the standard electric field strength for a $p$-brane source. We will also often use the magnetic formulation in which we have a non-zero $F_{8-p}$. The duality relation is
\begin{equation}
\star_{10} \sigma \left( F_{8-p} \right) = \e^{(3-p)\phi/2}F_{p+2}\,.
\end{equation}
The combination of $H_3$ and $F_{6-p}$ also induces $p$-brane charges dissolved in fluxes. We take these fluxes to be closed, i.e., $\d H_3 = 0 = \d F_{6-p}$. As usual, there is a subtlety with the self-dual 5-form field strength $F_5$. For $p=3$, the equations below have to be interpreted such that $F_{8-p}$ denotes the magnetic piece of $F_5$ and $F_{p+2}$ the electric part.

Our Ansatz for the form fields is
\begin{align}
& H_3= \d B_2 \,,\label{FormAnsatz1}\\
& C_{p+1}= \tilde \star_{p+1} \alpha ,\label{FormAnsatz2}\\
& \e^{- \phi} \star_{10} H_3 = \sigma(F_{6-p}) \w C_{p+1} - \tilde{\star}_{p+1} 1 \w X_{6-p}\,,\label{FormAnsatz3}
\end{align}
where $X_{6-p}$ is a closed $(6-p)$-form not proportional to $F_{6-p}$.\footnote{Any part of $X_{6-p}$ that is proportional to $F_{6-p}$ can be absorbed by a gauge transformation of $C_{p+1}$. This also implies that $X_{6-p}$ vanishes when $p=6$.}
The operator $\tilde \star_{p+1}$ is the Hodge star on the brane worldvolume metric with the warp factor $\e^{2A}$ and blackening factor $\e^{2f}$ excluded (i.e., the Hodge star associated to $\tilde{g}_{\mu\nu}$).
Due to the absence of transgression terms, we have $F_{p+2}= \d C_{p+1}$, which implies that the $H_3$ equation of motion
\begin{equation}
\d (\e^{-\phi}\star_{10} H_3) = F_{p+2}\wedge \sigma \left( F_{6-p} \right)
\end{equation}
is satisfied for \eqref{FormAnsatz3} as long as $F_{6-p}$ and $X_{6-p}$ are closed. The above Ansatz is general enough to describe anti-branes in a number of different backgrounds. For $p=3$, this includes anti-D3-branes in KT, which we already discussed in section \ref{antikt}, and anti-D3-branes in KS, which we will analyse in more detail below. For $p\neq3$, the Ansatz also covers setups such as the anti-D6-brane model discussed in \cite{Blaback:2011nz, Blaback:2011pn,Bena:2012tx, Bena:2013hr}.

We will now consider the remaining equations of motion for the above Ansatz and combine them in a such way that we obtain an equation analogous to (\ref{mastereq1}), but this time for non-compact spacetimes.
The trace-reversed Einstein equation along the brane worldvolume and the dilaton equation of motion are  
\begin{align}
& \tfrac{4}{p+1}R_{p+1} = - \tfrac{1}{2} \e^{- \phi} |H_3|^2 + \tfrac{p-7}{4} \e^{\tfrac{p-3}{2}\phi} |F_{8-p}|^2 + \tfrac{p-5}{4} \e^{\tfrac{p-1}{2} \phi} |F_{6-p}|^2 + \tfrac{p-7}{4} \mu_{p} \e^{\tfrac{p-3}{4}\phi} \delta \left( \Sigma \right)\,, \label{trE}\\
&0=- \nabla^2 \phi - \tfrac{1}{2} \e^{- \phi} |H_3|^2 + \tfrac{p-3}{4} \e^{\tfrac{p-3}{2}\phi} |F_{8-p}|^2 + \tfrac{p-1}{4} \e^{\tfrac{p-1}{2} \phi} |F_{6-p}|^2 + \tfrac{p-3}{4} \mu_{p} \e^{\tfrac{p-3}{4}\phi} \delta \left( \Sigma \right)\,, \label{dilaton}
\end{align}
where $\delta \left( \Sigma \right)$ is a delta distribution with support on the anti-brane worldvolume.
The combination (\ref{trE})  $ - \frac{p-5}{p-1} $ (\ref{dilaton}) gives
\begin{equation} \label{ME1}
\tfrac{4}{p+1} R_{p+1} =  \tfrac{p-5}{p-1} \nabla^2 \phi - \tfrac{2}{p-1} \left[\e^{-\phi}|H_3|^2  +\e^{\tfrac{p-3}{2}\phi} |F_{8-p}|^2 \right] - \tfrac{2}{p-1} \mu_{p} \e^{\tfrac{p-3}{4}\phi} \delta \left( \Sigma \right)\, .
\end{equation}
The left-hand side of this equation can be rewritten as
\begin{equation}\label{curv}
\tfrac{4}{p+1} R_{p+1} = \tfrac{4}{p+1} \e^{-2A}\tilde{R}_{p+1} -  \tfrac{4}{p+1} \e^{-(p+1)A-f} \nabla_m \partial^{m} \e^{(p+1)A+f} \, ,
\end{equation}
where $m$ runs over the transversal indices only.

We now aim at rewriting the right-hand side of (\ref{ME1}) in such a way that more terms combine into a total derivative. To this end, we use the Bianchi identity
\begin{equation}
\d F_{8-p} = H_3\wedge F_{6-p} - \mu_p\delta_{9-p}\,.
\end{equation}
After wedging from the left with $\sigma(C_{p+1})$, we can derive 
\begin{equation}
\begin{split}\label{eq:step1}
0 &= -(-1)^{p+1} \left[ \e^{-\phi} |H_3|^2 + \e^{(p-3)\phi/2} |F_{8-p}|^2\right] \star_{10} 1\\
&\quad +(-1)^{p+1} \d \left[ \sigma(C_{p+1})\w F_{8-p} + \tilde{\star}_{p+1} 1 \w B_2 \w X_{6-p}\right]\\
&\quad -(-1)^{p+1} \mu_p C_{p+1}\w \sigma(\delta_{9-p})\,.
\end{split}
\end{equation}
With this, (\ref{ME1}) takes the form
\begin{align}
\frac{4}{p+1} R_{p+1} \star_{10} 1 &= \frac{p-5}{p-1} \nabla^2 \phi \star_{10} + \frac{2}{p-1} \mu_p \left[ - \e^{(p-3)\phi/2} + \alpha \e^{-(p+1)A - f}\right]\star_{p+1} 1\w \sigma(\delta_{9-p}) \nonumber\\
& -\frac{2}{p-1}\d \left[ \sigma(C_{p+1})\w F_{8-p} + \tilde{\star}_{p+1} 1 \w B_2 \w X_{6-p} \right]\,. \label{ME2}
\end{align}
This result is consistent with \cite{Gautason:2013zw} for the choice $
c = - \frac{2}{p-1}$. Together with equation (\ref{curv}), this expression can be regrouped as follows:
\begin{align} \label{meq}
\frac{4}{p+1}\star_{10} \e^{-2A} \tilde{R}_{p+1} =&\, \frac{2}{p-1} \mu_p \left[ - \e^{(p-3)\phi/2} + \alpha \e^{-(p+1)A - f}\right]\star_{p+1} 1\w \sigma(\delta_{9-p})\nonumber\\
& + \d \left[ -\frac{2}{p-1}\left[ \sigma(C_{p+1})\w F_{8-p} +\tilde{\star}_{p+1} 1 \w B_2 \w X_{6-p} \right] \right.\nonumber \\
&\left. - \frac{p-5}{p-1} \star_{10} \d \phi - \frac{4}{p+1} \star_{10} \d \left[ (p+1)A+f \right ]\right]\,.
\end{align}
Except for the source terms and the cosmological constant term, all terms combine into a total derivative. As we will see below, it is necessary for our no-go theorem that this is the case. 

On a compact space and at zero $T$, the integrated version of \eqref{meq} simply reproduces the existing no-go theorem in \cite{Gautason:2013zw} that we reviewed in section \ref{review} (see also \cite{Burgess:2011rv}).
However, we now want to consider setups where the space transverse to the anti-D$p$-branes is non-compact. This can be done by integrating \eqref{meq} up to a boundary $\partial \mathcal{M}$. Together with $\tilde R_{p+1}= 2\frac{p+1}{p-1}\Lambda$, this yields the following extension of the master equation (\ref{mastereq1}):
\begin{equation}\label{mastereq}
\frac{8 v \mathcal{V}_{\scriptscriptstyle{\partial \mathcal{M}}}}{p-1} \Lambda =  \frac{2}{p-1} \left[ S^{(p)}_\textrm{DBI} + S_\textrm{WZ}^{(p)}\right] + \oint_{\partial \mathcal{M}} \mathcal{B}\,,
\end{equation}
where $\mathcal{V}_{\scriptscriptstyle{\partial \mathcal{M}}} = \int \star_{9-p}\, \e^{\left( p -1 \right) A+f}$ is the warped volume of the transverse space inside the boundary $\partial \mathcal{M}$. $\oint_{\partial \mathcal{M}} \mathcal{B}$ is a boundary term from integrating the total derivative in \eqref{meq} whose integrand is given by
\begin{align} \label{integrand}
\mathcal{B} =& -\frac{2}{p-1}\left[ \sigma(C_{p+1})\w F_{8-p} +\tilde{\star}_{p+1} 1 \w B_2 \w X_{6-p} \right] - \frac{p-5}{p-1} \star_{10} \d \phi \nonumber \\
& - \frac{4}{p+1} \star_{10} \d \left[ (p+1)A+f \right ].
\end{align}
All other objects in \eqref{mastereq} are defined as in section \ref{review}.

At $T=0$, we will choose the boundary $\partial \mathcal{M}$ to be located far away from the anti-brane position such that the integral over the transverse space runs from the anti-brane position up to a cutoff in the UV. At finite temperature, however, we encounter a small subtlety. In that case, we will argue below that the $H_3$ energy density blows up at the horizon, analogous to what we have already shown in section \ref{antikt} for the special case of anti-D3-branes in KT.
The easiest way to analyse this situation in our framework is then to choose an integration range with two boundaries, where we put the first one far away from the anti-branes and the second one very close to (but before) the horizon. The boundary term in \eqref{mastereq} then consists of two disconnected pieces with $\partial \mathcal{M} = \partial \mathcal{M}_\text{horizon} \cup\partial \mathcal{M}_\text{UV}$.

\subsection{The no-go theorem at $T=0$}
\label{nogo-zeroT}

Similar to the argument reviewed in section \ref{review}, we will now use the master equation \eqref{mastereq} to show that anti-branes generate a singularity in the $H_3$ energy density in the non-compact backgrounds we consider. In the following, we will restrict to $\Lambda=0$. This is sufficient to discuss anti-D3-branes in KS, which is the setup we are mainly interested in. We furthermore specialise to the case $T=0$ for the moment, i.e., we set $\e^{2f}=1$. The extension of our arguments to finite temperature and non-zero cosmological constant will be discussed in sections \ref{nogo-finiteT} and \ref{nogo-finiteLambda}.

For $\Lambda=0$ and $T=0$, it is now easy to follow the logic of section \ref{review}. \eqref{mastereq} then relates the boundary conditions of the fields at the anti-brane position to a boundary term that only depends on the field values far away from the anti-branes,
\begin{equation}\label{mastereq0}
\frac{2}{1-p} \left[ S^{(p)}_\textrm{DBI} + S_\textrm{WZ}^{(p)}\right] = \oint_{\partial \mathcal{M}} \mathcal{B}\,.
\end{equation}
Using \eqref{braneaction} and \eqref{FormAnsatz2} and assuming the standard behaviour $\e^{(p+1)A+\frac{p-3}{4}\phi} \to 0$ near the anti-branes, it then follows
\begin{equation}\label{alpha0-boundary}
\alpha_0 = \frac{p-1}{2v{\bar N}\mu_p}\, \oint_{\partial \mathcal{M}} \mathcal{B},
\end{equation}
where ${\bar N}$ is the anti-brane number and $\alpha_0$ denotes the value of the gauge potential profile $\alpha$ at the position of the anti-branes.

Hence, if the boundary term $\oint_{\partial \mathcal{M}} \mathcal{B}$ is non-zero in a given solution with $p\neq1$, also $\alpha$ is non-zero at the anti-brane position. The $H_3$ energy density then contains terms that diverge at least as bad as
\begin{equation}
\e^{-\phi}|H_3|^2\sim \e^{-2(p+1)A + \phi} \alpha^2 |F_{6-p}|^2 \sim \e^{-2A}
\end{equation}
in the vicinity of the anti-branes. Here, we assumed that the dilaton and the internal metric diverge such that they locally take the form $\e^{\phi} = \e^{4\frac{p-3}{7-p}A}\e^{\tilde \phi}$, $g_{mn} = \e^{2\frac{p+1}{p-7}A} \tilde g_{mn}$ close to the anti-branes (with $\e^{\tilde \phi}$ and $\tilde g_{mn}$ regular), as it is the case for the standard D-brane solutions in flat space.
Hence,
\begin{equation}
\oint_{\partial \mathcal{M}} \mathcal{B}\neq 0\qquad \rightarrow \qquad\text{singularity}\,.
\end{equation}

This leaves us to establish that the boundary term is indeed non-zero.
First of all, since the source terms are independent of the position of the boundary $\partial \mathcal{M}$ in \eqref{mastereq0}, it follows that also the value of the boundary term itself must be independent of its position. Hence, $\oint_{\partial \mathcal{M}} \mathcal{B}$ is a conserved charge.

In appendices \ref{Secbandm} and \ref{SecEvalbandm}, we have computed the value of the boundary term for anti-D3-branes in the KS throat. This setup is captured by the above equations for the choice $p=3$.
In order to evaluate the integrand \eqref{integrand}, we have used the explicit solution found in \cite{Bena:2009xk}, which is valid at linear order in the number of anti-branes ${\bar N}$ and in the approximation where the branes are partially smeared over the tip of the deformed conifold. However, if we consider a boundary in the UV that is far enough away from the tip, corrections due to the localisation of the anti-branes should be negligible in \eqref{integrand}. Furthermore, if we choose the number of anti-branes ${\bar N}$ to be small compared to the units of D3-brane charge induced by the fluxes, non-linear corrections in ${\bar N}$ are expected to be suppressed in the UV as well. Substituting the solution of \cite{Bena:2009xk} into our boundary term, we find that, up to a volume factor, it equals the ADM mass of the solution, which was computed in \cite{Dymarsky:2011pm}. Hence,
\begin{equation}\label{result2}
\boxed{\frac{1}{v}\oint_{\partial \mathcal{M}} \mathcal{B} = M =  2 \e^{4A_\text{BPS}} \bar N \mu_{3}\,}
\end{equation}
at first order in the number of anti-branes, where $\e^{4A_\text{BPS}}$ denotes the (finite) value of the warp factor at the tip of the unperturbed KS background (i.e., the BPS solution without any anti-D3-branes).\footnote{Here, $M$ refers to the ADM mass normalised w.r.t. the BPS background.}
Using this in \eqref{alpha0-boundary}, we observe that the ADM mass fixes the IR boundary condition of $\alpha$ to
\begin{equation}
\alpha_0 = 2 \e^{4A_\text{BPS}}.
\end{equation}
Note that this confirms a conjecture made previously in \cite{Junghans:2014xfa} from the point of view of a 4D warped effective field theory analysis.

Under the assumptions discussed above, this then implies a singularity in the $H_3$ energy density. We should stress here that this conclusion does not only hold for the linearised, partially smeared solution of \cite{Bena:2009xk}. Instead, the master equation \eqref{mastereq0}, which relates the boundary term to $\alpha_0$ and thus leads to the singularity, is valid for fully localised and non-linearly backreacting anti-branes. We have only used the explicit solution of \cite{Bena:2009xk} to verify that the boundary term is indeed non-zero in the solution with anti-branes. Since we can put the boundary arbitrarily far away from the anti-branes, corrections due to the linearisation or the partial smearing are expected to be small in $\oint_{\partial \mathcal{M}} \mathcal{B}$, as we have argued above. We therefore expect that the boundary term is non-zero also in the non-linear and fully localised anti-brane solution.

Moreover, we expect that this correspondence between the boundary term and the ADM mass is not a coincidence but also holds for the other solutions in the class described by our Ansatz. First, as discussed above, the fact that $\oint_{\partial \mathcal{M}} \mathcal{B}$ is constant for $\Lambda = 0$ suggests that it equals a charge. Since that charge is the ADM mass for a specific example (i.e., KS), it is most likely related to the ADM mass for all examples. We have checked that this is indeed the case for the (non-compact) anti-D6-brane models discussed in \cite{Blaback:2011nz, Blaback:2011pn, Bena:2012tx, Bena:2013hr}.
Second, to compute the mass (energy) of a spacetime, one should use the ADM formalism, which uses the bulk action and leaves out the explicit source actions. The energy of that source can also be measured using the source action alone. This is a well-known result in case backreaction can be ignored, because then one also finds
\begin{equation}
M =  2 \e^{4A_\text{BPS}} \bar N \mu_{3}\,
\end{equation}
by simply summing the DBI and WZ action of the probe. What our above result (\ref{mastereq}, \ref{result2}) shows is that the on-shell source action equals the ADM mass even when it is evaluated using the backreacted solution. In that case, the DBI term vanishes since the warp factor vanishes at the anti-brane but the WZ term gives a finite contribution since the gauge potential $\alpha$ is non-zero (in the gauge we fixed). Its non-zero value $\alpha_0$ is then exactly such as to give the ADM mass.

\subsection{The no-go theorem at $T \neq 0$}
\label{nogo-finiteT}

Let us now discuss the extension of the above arguments to the case of finite temperature. Analogous to the $T=0$ case, \eqref{mastereq} then implies that the 3-form energy density has singular terms at the horizon if the boundary term $\oint_{\partial \mathcal{M}} \mathcal{B}$ and the fields at the horizon satisfy certain conditions. Let us at first list these conditions in detail:
\begin{enumerate}
 \item Locally near the horizon, the metric coordinates can be split into coordinates along the horizon and a normal coordinate $r$.\footnote{The normal coordinate $r$ is not necessarily the radial coordinate in the transverse space. For partially smeared anti-D3-branes in KS, for example, the horizon is expected to be a surface of constant $\tau$ \cite{Bena:2012ek}.} The horizon $r=r_0$ is characterised by a coordinate singularity in the time and normal components of the metric with $g_{tt} = -\e^{2f}$, $g_{rr} = \e^{-2f}\tilde g_{rr}$ and $\e^{2f} \sim r-r_0$ (at leading order) whereas all other fields remain finite. 
 \item The variation of the fields $A$, $f$ and $\phi$ along the horizon is negligible, i.e., $A=A(r)$, $f=f(r)$, $\phi=\phi(r)$ near the horizon.
 \item The form fields and the metric satisfy the Ansatz described in section \ref{section-mastereq}.
 \item The component of $F_{6-p}$ with all legs along the horizon is non-vanishing.
 \item The boundary term satisfies
 \begin{equation}\label{cond-boundary}
 \frac{1}{v}\oint_{\partial \mathcal{M}_\text{UV}} \mathcal{B} = c(M_0,T) - \frac{2}{p+1}TS,
 \end{equation}
where $M_0$ is the value of the ADM mass at $T=0$, $S$ is the entropy associated to the black hole horizon and $c(M_0,T)$ is an analytic function of $T$, which is identically zero for $M_0=0$ but non-trivial otherwise.
\end{enumerate}
The first two assumptions are motivated by the usual behaviour of black holes and black branes, while assumptions three and four are due to the specific setups we are interested in, i.e., anti-D3-branes in KT and KS as well as their $(p+1)$-dimensional generalisations. The last assumption will be discussed below in more detail, where we will argue that it is the natural generalisation of our previous result that the boundary term is proportional to the ADM mass at $T=0$.

Under the above conditions, we can derive a no-go theorem from \eqref{mastereq} as follows. As discussed above, we now choose the integration space in \eqref{mastereq} such that it has two boundaries, one far away from the anti-branes and the second one near the horizon. Since there are no source terms in the enclosed region, \eqref{mastereq} then simply equates the boundary term at the horizon to the one in the UV,
\begin{equation}\label{mastereq00}
\oint_{\partial \mathcal{M}_\text{horizon}} \mathcal{B} = \oint_{\partial \mathcal{M}_\text{UV}} \mathcal{B}.
\end{equation}
With $\e^f \to 0$ and $A=A(r), \phi=\phi(r)$, the integrand \eqref{integrand} of the boundary term at the horizon reduces to
\begin{equation} \label{integrand00}
\mathcal{B}|_\text{horizon} = -\frac{2}{p-1} \tilde{\star}_{p+1}1 \w \left[ \alpha \sigma(F_{8-p}) + B_2 \w X_{6-p} \right] - \frac{4}{p+1} \star_{10} \d f,
\end{equation}
where we also used that $C_{p+1} = \tilde \star_{p+1} \alpha$.

The last term on the right-hand side of the above equation is determined by the behaviour of the blackening factor at the horizon and is expected to be proportional to the temperature and the entropy of the black hole.
This can be verified for a general black hole using the standard Euclidean path integral formalism, with the result
\begin{equation}\label{dfterm}
- \frac{1}{v} \oint_{\partial \mathcal{M}_\text{horizon}} \star_{10} \d f = -\frac{1}{2}TS.
\end{equation}
Note that this follows only from the local form of $f(r)$ near the horizon and is independent of whether the black hole carries anti-D$p$-brane charge or not.
Together with \eqref{cond-boundary}--\eqref{integrand00}, one then finds
\begin{equation}
- \frac{1}{v}  \frac{2}{p-1} \oint_{\partial \mathcal{M}_\text{horizon}} \tilde{\star}_{p+1}1 \w \left[ \alpha \sigma(F_{8-p}) + B_2 \w X_{6-p} \right] = c(M_0,T).
\end{equation}
Hence, whenever $c(M_0,T)$ is non-zero, $\alpha$ and $X_{6-p}$ cannot vanish simultaneously at the horizon.
This then implies that singular terms appear in the $H_3$ energy density, which diverge at least as bad as
\begin{equation}\label{fgsghl}
\e^{-\phi}|H_3|^2 \sim \e^{-2(p+1)A - 2f + \phi} |\alpha F_{6-p} + X_{6-p}|^2 \sim \e^{-2f},
\end{equation}
where we used the assumption that $F_{6-p}$ has a non-vanishing component with all legs along the horizon and that the metric diverges as stated above.
For $p=3$, this reproduces the singular density (\ref{KTfiniteT}) we derived for anti-D3-branes in the KT background using a different technique. The advantage of the present formulation is that it is valid for a more general class of setups, including anti-D3-branes in KS.

A crucial step in the above argument is assumption \eqref{cond-boundary} for the general form of the boundary term at finite $T$. Let us therefore motivate this assumption by comparing our general equations with explicit solutions in the literature. For concreteness, we will again focus on the special case $p=3$ as in the previous section, i.e., anti-D3-branes in the KT or KS background. In \cite{Aharony:2007vg}, the authors constructed black hole solutions with positive D3-brane charge in the KT background. In the limit $T\to 0$, these solutions reduce to BPS solutions with extremal D3-branes (and no anti-D3-branes) and therefore have $M_0=0$. Substituting the ansatz of \cite{Aharony:2007vg} into \eqref{integrand}, we find\footnote{It was observed in \cite{Aharony:2007vg} that $TS$ is a renormalisation group flow invariant in their class of solutions, i.e., it is independent of the radial coordinate. This fits nicely together with our earlier remark that the value of the boundary term for $\Lambda=0$ is independent of the position of the boundary.}
\begin{equation}\label{boundary-finiteT}
\frac{1}{v}\oint_{\partial \mathcal{M}_\text{horizon}} \mathcal{B} = \frac{1}{v}\oint_{\partial \mathcal{M}_\text{UV}} \mathcal{B} = -\frac{1}{2}TS.
\end{equation}
One can furthermore verify that \eqref{dfterm} is satisfied in these solutions. Together with our earlier result \eqref{result2} for anti-D3-branes in KS at $T=0$, this is sufficient to infer some general statements about the boundary term for arbitrary $M_0$ and $T$.

First, one might have guessed that a natural generalisation of the results of section \ref{nogo-zeroT} is that, at finite $T$, the boundary term is related to a thermodynamical observable such as the internal energy or the free energy. The above expression shows that this is not the case: neither the internal energy nor the free energy are proportional to $TS$ in the solutions of \cite{Aharony:2007vg}.
Second, comparing \eqref{boundary-finiteT}, which is valid for $M_0=0,T\neq0$, with \eqref{result2}, which is valid for $M_0\neq 0,T=0$, it is straightforward to write down a general ansatz for the boundary term for arbitrary $M_0$ and $T$. Assuming that the form of the boundary term is universal, we can write
\begin{equation}\label{result3}
\frac{1}{v}\oint_{\partial \mathcal{M}_\text{UV}} \mathcal{B} = c(M_0,T) - \frac{1}{2}TS
\end{equation}
with $c(0,T)=0$ and $c(M_0,0)=M_0$. This indeed agrees with condition \eqref{cond-boundary} for the special case $p=3$.

An important question is whether, for general $M_0$ and $T$, the function $c(M_0,T)$ could be such that the above no-go theorem is evaded when enough temperature is turned on. This would require that $c(M_0,T)$ is zero above some critical temperature $T_\text{crit}$ for all temperatures $T>T_\text{crit}$. However, if $c(M_0,T)$ is analytic in $T$, this can only happen if it is identically zero. Since $c(M_0,0)=M_0$, this is not the case for all solutions with $M_0 \neq 0$, i.e., those that carry anti-D3-brane charge at the horizon. Accordingly, the flux singularities of these solutions are not cloaked at any temperature. This conclusion is independent of the explicit form of $c(M_0,T)$ in terms of $M_0$ and $T$.\footnote{As an illustration, consider the simple example $c(M_0,T) = M_0 \left( 1- T/T_\text{crit} \right)$. This is zero at $T=T_\text{crit}$ such that the no-go theorem would be evaded at that temperature. However, the singularity would reappear for any temperature $T > T_\text{crit}$ and is therefore not cloaked by the black hole horizon.} For black hole solutions with positive D3-brane charge, on the other hand, we have $M_0=0$. Since $c(0,T)$ is identically zero, we do then not expect any flux singularities to appear. Indeed, one can explicitly check that the $H_3$ energy density is regular at the horizon in the solutions of \cite{Aharony:2007vg}.

A possible way to evade our no-go theorem would be to relax the assumption that $c(M_0,T)$ is analytic in $T$. One can then easily find examples of piecewise-defined functions that are non-zero below $T_\text{crit}$ but zero for all higher temperatures such that the flux singularity would be cloaked for $T>T_\text{crit}$. It is not clear to us whether such a behaviour is physically reasonable since $c(M_0,T)$ can be obtained by integrating fields over a UV boundary far away from the horizon. On the other hand, we cannot exclude that, e.g., a phase transition could yield a non-analytic $c(M_0,T)$. We leave a detailed analysis of this question for future work. Let us note, however, that our alternative no-go in section \ref{antikt} for anti-D3-branes in KT and the numerical results of \cite{Bena:2012ek} for anti-D3-branes in KS and KT suggest that flux singularities are not resolved this way.

As we pointed out in section \ref{review}, another caveat is that several singular terms in \eqref{fgsghl} might conspire to cancel each other out at the horizon. In contrast to the $T=0$ case, there are not many explicit results available for anti-brane solutions at finite $T$ such that it is difficult to judge whether such a cancellation should be expected or not. In the numerical studies of \cite{Bena:2012ek}, this was not the case for partially smeared anti-D3-branes in KS, and the $H_3$ energy density was indeed found to be singular at the horizon. While our analytic arguments are in agreement with this result, we cannot exclude that configurations could exist in which a resolution of the singularity happens at finite $T$ via such a mechanism.

\subsection{The no-go theorem for $\Lambda\neq 0$}
\label{nogo-finiteLambda}

Finally, we consider what happens if $\Lambda \neq 0$, where we specialise to the case $T=0$ for simplicity. The reasoning here is slightly different from what was done for compact models at $T=0$ in section \ref{review}. In a non-compact model, the $(p+1)$-dimensional cosmological constant is not constrained by the integrated Einstein equations. Contrary to a compact setting, it is therefore not necessarily related to the energy density of uplifting terms such as anti-branes but simply a parameter that can be chosen freely.

Adding a cosmological constant term to the equations of motion yields an extra amount of energy density, which is distributed over the whole transverse space (weighted with a warp factor). The master equation (\ref{mastereq}) then receives an extra contribution from the total energy due to $\Lambda$, which depends on the integration volume. Thus, the value of the boundary term is not independent of the position of the boundary anymore. It follows from (\ref{mastereq}), however, that the boundary term and the $\Lambda$ term taken together are still a conserved quantity, which suggests that they now measure the contribution of the anti-brane to the total energy. We leave an explicit derivation of the boundary term for setups with non-zero $\Lambda$ for future work. We can however conclude that the absence of the singularity requires a precise cancellation between the boundary term and the cosmological constant in (\ref{mastereq}). Our above interpretation suggests that such a cancellation does not happen in the presence of anti-branes at the tip. This is in line with numerical results of \cite{Buchel:2013dla}.

\section{Discussion}
\label{discussion}

The main result of our paper can be summarised by equations (\ref{mastereq}, \ref{result2}) and (\ref{result3}). These equations demonstrate that, under certain assumptions, anti-branes in warped throats necessarily have a singular 3-form energy density at both zero and finite temperature. This result did not rely on numerics, on smearing branes or on using a specific geometry. It is therefore important to highlight the assumptions we did use.

First, we have assumed an Ansatz \eqref{metricANSATZ}, (\ref{FormAnsatz1}), (\ref{FormAnsatz2}), (\ref{FormAnsatz3}) for the metric and the form fields that captures a large class of setups including anti-D3-branes in KS but might be evaded by more complicated flux backgrounds that do not fall into this class (e.g., due to the presence of additional topological fluxes). We also made assumptions on the near-brane behaviour of some of the fields that are well-motivated by known brane and black hole solutions but might in principle be evaded.

Second, our no-go theorem requires that the boundary term $\oint_{\partial \mathcal{M}} \mathcal{B}$ is non-zero.
One of the key results of this paper is that the boundary term equals the ADM mass at $T=0$ for the case of anti-branes in the KS geometry. This leads us to conjecture that, at finite $T$, the boundary term is given by the general expression \eqref{result3}, where we leave an explicit derivation of the function $c(M_0,T)$ for future work.

Apart from the above caveats, our arguments rule out the existence of regular solutions for anti-D3-branes in KT and anti-D3-branes smeared over the tip in KS. Hence, our analytic results confirm the numerical results in \cite{Bena:2012ek}. As explained above, it is furthermore reasonable to assume that our Ansatz captures localised anti-branes in KS as well (see also \cite{Gautason:2013zw}).
Our results thus have three possible interpretations: either 1) the Gubser criterium \cite{Gubser:2000nd} fails, or 2) the anti-brane backgrounds are pathological and one does not expect a resolution to occur, or 3) the resolved solutions evade one of our assumptions.

\section*{Acknowledgements}
We would like to thank Iosif Bena, Alejandra Castro and Anatoly Dymarsky for useful discussions. DJ would like to thank the University of Leuven for hospitality during a visit where part of this work was completed. The work of JB, UHD and SCV is supported by the Swedish Research Council (VR). TVR is supported by a Pegasus fellowship and by the Odysseus programme of the FWO.

\appendix

\section{Boundary term and ADM mass in KS} \label{Secbandm}

In what follows, we look in more detail at the solution of \cite{Bena:2009xk} describing partially smeared anti-D3-branes at $T=0$ at the tip of the KS throat. This example serves to illustrate some of the general statements we make in the main text regarding the boundary term when  $\Lambda=0$. 
\subsection{The boundary term}\label{ssec:appa1} 

The Ansatz is
\begin{eqnarray}\label{ansatz-smearedks}
&& \mathrm{d} s^2_{10} = \e^{2A+2p-x} \mathrm{d} s^2_{1,3} + \e^{-6p-x} \mathrm{d} \tau^2 + \e^{x+y} \left( g^2_1 + g^2_2 \right) + \e^{x-y} \left( g^2_3 + g^2_4 \right) + \e^{-6p-x} g^2_5\,,\nonumber
\\
&& H_3 = \frac{1}{2} \left( k-f \right) g_5 \wedge \left( g_1 \wedge g_3 + g_2\wedge g_4 \right)+\mathrm{d} \tau \wedge \left( f' g_1 \wedge g_2 + k' g_3\wedge g_4 \right)\,,\nonumber
\\
&& F_3 = F g_1 \wedge g_2 \wedge g_5 + \left( 2P - F \right) g_3 \wedge g_4 \wedge g_5 + F' \mathrm{d} \tau \wedge \left( g_1 \wedge g_3 + g_2 \wedge g_4 \right)\,,\nonumber
\\
&& \phi = \phi \left( \tau \right)\,,\nonumber
\\
&& C_0 = 0 \,,
\end{eqnarray}
where $A,p,x,y,k,f,F$ are functions of $\tau$ and $P$ is a constant. This Ansatz reduces to the KT Ansatz for the choice $y=0$, $F=P$ and $k=f$ \cite{Bena:2011wh}. We have borrowed the notation from \cite{Bena:2009xk}, which implies that the symbol $p$ does not denote the dimension of the brane anymore and $f$ does not relate to the blackening factor of equation (\ref{metricANSATZ}).
In addition, let us write (as in \cite{Gautason:2013zw})
\begin{eqnarray}
&& C_4 = \tilde \star_4 \alpha \,, \nonumber
\\
&& F_5 = - \left( 1 + \star_{10} \right) \e^{-4 \tilde A} \star_6 \mathrm{d} \alpha \,,
\end{eqnarray}
where we use $\tilde A$ to distinguish it from the function $A$. $\tilde A$ is precisely the function associated to the warp factor of the external coordinates: $
\e^{4 \tilde A} \tilde \star_{4} 1 = \star_{4} 1$, or
\begin{equation}
\tilde A = A + p - \frac{1}{2}x .
\end{equation}

We now follow (\ref{FormAnsatz3}) and write
\begin{equation}
H_3 = \e^{\phi - 4 \tilde A} \star_6 \left( \alpha F_3 + X_3 \right).
\end{equation}
The form $X_3$ can be found explicitly:
\begin{align}
& X_3 = \left[ \e^{- \phi + 4 A + 4 p - 2 x} \left( \frac{k-f}{2} \right)  - \alpha F' \right] \mathrm{d} \tau \wedge \left( g_1 \wedge g_3 +  g_2 \wedge g_4 \right) + \\ & \left[ \e^{- \phi + 4 A + 4 p - 2 x +2y} k' - \alpha F \right] g_1 \wedge g_2 \wedge g_5   + \left[ \e^{- \phi + 4 A + 4 p - 2 x - 2y } f' - \alpha \left( 2 P - F \right) \right] g_3 \wedge g_4 \wedge g_5\,.\nonumber
\end{align}
From the closure of $X_3$, we can derive that 
\begin{equation}
\alpha =  \lambda \e^{ 4 A + 4 p - 2 x} + a_0\,,\qquad\text{where}\qquad  \lambda = \frac{1}{2 P } \left(f' \e^{-2y} + k' \e^{2y} \right) \e^{- \phi}\,
\end{equation}
and $a_0$ is a constant, which we take zero from now on as a gauge choice. This can then be used to rewrite $X_3$ as
\begin{equation}
X_{3} = \mathrm{d} \omega_2 \,
\end{equation}
with
\begin{equation}
\omega_2 = \left[ \frac{1}{2} \left( k' \e^{2y} - f' \e^{-2y}\right) \e^{-\phi+4A+4p-2x} - \alpha F + P \alpha  \right] \left( g_1 \wedge g_3 + g_2 \wedge g_4 \right)\,.
\end{equation}

We also have
\begin{equation}
H_3 = \mathrm{d} \left[ f g_1 \wedge g_2 + k g_3 \wedge g_4 \right]\,
\end{equation}
and, as a consequence, $H_3 \wedge X_3 =  - \mathrm{d} \left[ \left(  \mathrm{d} B_2 \right) \wedge   \omega_2 \right]$.
This means that the contribution in the master equation $
- \tilde \star_{4} 1 \wedge H_{3} \wedge X_{3} = \mathrm{d} \left[  \tilde \star_{4} 1 \wedge \left(  \mathrm{d} B_2 \right) \wedge   \omega_2 \right]$
adds as a boundary term as in \eqref{meq}. The complete integrand in the boundary term is then
\begin{equation}
\d \mathcal{B}= \e^{-4A-4p+2x} \left( \nabla_m \partial^m \e^{4A+4p-2x} \right)  \star_{10} 1 - \nabla^2 \phi \star_{10} 1 - \mathrm{d} \left[  C_{4}  \wedge F_{5} \right] - \tilde \star_{4} 1 \wedge H_{3} \wedge X_{3}. 
\end{equation}
Writing the boundary term explicitly in our Ansatz, we find
\begin{equation}
\frac{1}{2\kappa_{10}^2} \oint \mathcal{B}= \frac{1}{2\kappa_{10}^2} \int   b\,\,\tilde \star_{4} 1 \wedge g_1 \wedge g_2 \wedge g_3 \wedge g_4 \wedge g_5 ,
\end{equation}
where we restored the constant $2\kappa_{10}^2$ and
\begin{eqnarray}
&& b =  \e^{4A+4p} \left(4A+4p-2x - \phi \right)' - \e^{- 4 A - 4 p + 4 x} \alpha' \alpha 
\\
&& \ \ \ \ \ -  \left( k - f \right) \left[ \frac{1}{2} \left( k' \e^{2y} - f' \e^{-2y}\right) \e^{-\phi+4A+4p-2x} - \alpha F + P \alpha   \right]  \nonumber .
\end{eqnarray}

\subsection{The ADM mass}

The ADM mass $M$ of the solution of \cite{Bena:2009xk} was computed in \cite{Dymarsky:2011pm}. The general expression in terms of the fields defined in \eqref{ansatz-smearedks} is
\begin{equation}
M = - \frac{48}{(2\pi)^4 \alpha'^4} m, \quad  m = \e^{4A+4p} A' - \frac{W}{3},
\end{equation}
where the superpotential
\begin{eqnarray}
W= \e^{4A-2p-2x}+\frac{1}{2}\e^{4A+4p+y}+\frac{1}{2}\e^{4A+4p-y}+\frac{1}{2}\e^{4A+4p-2x}\left[ f_0 \left( 2 P - F_0 \right) + k_0 F_0 \right]
\end{eqnarray}
acts as a normalisation of $M$ to provide a finite result.

\section{Evaluation of boundary term and ADM mass }\label{SecEvalbandm}

\subsection{The boundary term}

Let us consider a first order perturbation around the KS solution as in \cite{Bena:2009xk} and compare the boundary term and the ADM mass. All the following expressions should be understood to be valid up to $\mathcal{O} \left(\bar{N}^2 \right)$, where $\bar{N}$ is the number of anti-branes.
We then write
\begin{eqnarray}
&& A = A_0 + \bar{N} A_1 ,
\\
&& p = p_0 + \bar{N} p_1 ,
\\
&& x = x_0 + \bar{N} x_1 ,
\\
&& y = y_0 + \bar{N} y_1 ,
\\
&& k = k_0 + \bar{N} k_1 ,
\\
&& f = f_0 + \bar{N} f_1 ,
\\
&& F = F_0 + \bar{N} F_1 ,
\\
&& \phi = \phi_0 + \bar{N} \phi_1 .
\end{eqnarray}
Note that our notation is such that $\bar{N}$ appears explicitly in all the expressions. Similarly, we may write
\begin{eqnarray}
&& \lambda = \lambda_0 + \bar{N} \lambda_1 = -1 + \bar{N} \lambda_1\,,
\\
&& \alpha = \alpha_0 + \bar{N} \alpha_1\,,
\\
&& \e^{4 A + 4 p - 2 x } = - \alpha_0 + \bar{N} v^{4}_1\,,
\\
&&  \frac{1}{2} \left[ k' \e^{2y} - f' \e^{-2y} \right] = P - F_0 + \bar{N} \left[  \frac{1}{2} \left( k'_1 \e^{2y_0} - f'_1 \e^{-2y_0} \right) - 2 y_1 P \right] ,
\end{eqnarray}
where
\begin{eqnarray}
&& \lambda_1 = \phi_1 + \frac{1}{2P} \left[ f'_1 \e^{-2y_0} + k'_1 \e^{2y_0} \right] + 2 y_1 - \frac{2}{P} y_1 F_0\,,
\\
&& \alpha_1 = - \alpha_0 \lambda_1 + \alpha_0 \left( 4 A_1 + 4 p_1 - 2 x_1 \right)\,,
\\
&& v^{4}_1 =  - \alpha_0 \left( 4 A_1 + 4 p_1 - 2 x_1 \right) = - \alpha_1 - \alpha_0 \lambda_1 .
\end{eqnarray}
To proceed, we make use of some zeroth-order relations that are given in appendix \ref{AppConventions}. Up to this order, one finds
\begin{eqnarray}
&& b= \bar{N}  b_1 ,
\\
&& m = \bar{N} m_1 ,
\end{eqnarray}
with
\begin{eqnarray}
&& \frac{b_1}{\alpha_0 } =  2 \lambda_1 \left[ f_0 \left( 2 P - F_0 \right) + k_0 F_0 \right]  + \e^{2 x_0} \left(\phi'_1 -  \lambda'_1\right) \nonumber
\\
&& \ \ \ \ \ \ \ \  - \left( k_0 - f_0 \right) \left[ F_0 \left( \lambda_1 - \phi_1 + 2 y_1 \right) - F_1 - k'_1 \e^{2 y_0} \right] ,
\\
&& \frac{m_1}{\alpha_0 } = - 2 x_1 \e^{2 x_0} A'_0 - \e^{2 x_0} A'_1 - 2 p_1 \e^{- 6 p_0} + \frac{1}{6} \left(2 x_1 + y_1 \right) \e^{2 x_0 + y_0} \nonumber
\\
&& \ \ \ \ \ \ \ \ + \frac{1}{6} \left(2 x_1 - y_1 \right) \e^{2 x_0 - y_0} + \frac{1}{6} \left[ f_1 \left( 2 P - F_0 \right) - f_0 F_1 + k_0 F_1 + k_1 F_0 \right] .
\end{eqnarray}
Both of these expressions can be further simplified. The relations in the appendices A and B of \cite{Bena:2011wh} provide us with explicit expressions for the derivatives of first order corrections to the functions. Again, one important remark about conventions is that in \cite{Bena:2011wh} (as in most of the previous work) the first order variations of the functions are defined by including the factor of $\bar{N}$. For instance, the functions $\tilde \phi^\prime_i$ in equations (113), (117) and (118) of \cite{Bena:2011wh} correspond to
\begin{eqnarray}
&& \tilde  \phi'_8 = \bar{N} \phi'_1\,,
\\
&&\tilde  \phi'_5 = \bar{N} f'_1\,,
\\ 
&& \tilde \phi'_6 = \bar{N} k'_1\,
\end{eqnarray}
in terms of our definition for the first order perturbations. With these equations, we find
\begin{eqnarray}
&& \bar{N} \lambda_1 = \frac{2}{P \alpha_0} \tilde \xi_5\,,
\\
&& \bar{N} k'_1 \e^{2 y_0} = 2 y_1 F_0 - F_0 \phi_1 - F_1 + \frac{2}{\alpha_0} \left( \tilde \xi_5 - \tilde \xi_6 \right) \,,
\\
&& \bar{N} \phi'_1 = 4 \frac{\e^{-2 x_0}}{\alpha_0} \tilde \xi_8 .
\end{eqnarray}
In addition, we make use of equations (107) and (121) in \cite{Bena:2011wh} to write
\begin{equation}
\tilde \xi'_5 = - \frac{1}{3} P e ^{-2 x_0} h(\tau) X_1 ,
\end{equation}
where $ h(\tau) = \e^{-4 A_0 - 4 p_0 + 2 x_0 }$. Putting all this together, we find for the boundary term
\begin{eqnarray}
&& \bar{N} b_1 = \frac{2}{P} \left[ f_0 \left( 2 P - F_0 \right) + k_0 F_0 \right]  \tilde \xi_5 + 4 \tilde \xi_8 + \frac{2}{3} X_1 h(\tau) + 2 \left( k_0 - f_0 \right) \left[ \tilde \xi_5 \left( 1 - \frac{F_0}{P} \right) - \tilde \xi_6 \right] \nonumber
\\
&& \ \ \ \ \ \ = 4 \tilde \xi_8 + 2 \left( k_0 + f_0 \right) \tilde \xi_5 + 2 \left( f_0 - k_0 \right) \tilde \xi_6 + \frac{2}{3} X_1 h(\tau).
\end{eqnarray}
With equation (31) in \cite{Bena:2011hz} for $\tilde \xi_8$, we find
\begin{equation}
\bar{N} b_1 = 4 X_8 = X_4 = \frac{\pi}{2 h(0)} \bar{N} .
\end{equation}
Using $2 \kappa^2_{10} = (2\pi)^7 \alpha'^4$ (the constant $\alpha'$ should not be confused with $\partial_\tau \alpha$) and
\begin{eqnarray}
&& \int g_1 \wedge g_2 \wedge g_3 \wedge g_4 \wedge g_5 = 64 \pi^3\,,
\\
&& \int \tilde \star_{4} 1 \equiv v ,
\end{eqnarray}
we finally find
\begin{equation}\label{BTERM}
\frac{1}{2 \kappa^2_{10}}   \oint \mathcal{B} =  \frac{8}{(2\pi)^4 \alpha'^4} v X_4 + \mathcal{O}  \left(\bar{N}^2 \right).
\end{equation}

\subsection{The ADM mass}

We can use the same treatment for the ADM mass at first order in $\bar{N}$. Equations (116) and (120) in \cite{Bena:2011wh} give
\begin{eqnarray}
 && \bar{N} \e^{2 x_0} A'_1 = \frac{\e^{2 x_0}}{3} \left( \tilde \phi'_4 - \tilde \phi'_1 \right)  = \frac{X_4}{6 \alpha_0} - 2 \e^{ - 6 p_0 } p_1 + \frac{1}{6} \left( \e^{2 x_0 + y_0 } - \e^{2 x_0 - y_0} \right) y_1 + \\
&& \ \ \ \ \ \   \frac{1}{6} \left[ f_1 \left( 2 P - F_0 \right) - f_0 F_1 + k_0 F_1 + k_1 F_0 \right] + 2 x_1 \left\{-\frac{1}{6} \left[ f_0 \left( 2 P - F_0 \right) + k_0 F_0 \right] -\frac{1}{3} \e^{-6 p_0} \right\} \nonumber .
\end{eqnarray}
With this, we find for the ADM mass
\begin{equation}
\bar{N} m_1 =  - \frac{1}{6} X_4 - 2 \alpha_0 x_1 \left\{ \e^{2x_0} A'_0 - \frac{1}{3} \e^{-6p_0}- \frac{1}{6}\e^{2 x_0 + y_0}- \frac{1}{6}\e^{2x_0 - y_0 }- \frac{1}{6}\left[ f_0 \left( 2 P - F_0 \right) + k_0 F_0 \right] \right\} ,
\end{equation}
which, using the zeroth-order information, becomes
\begin{equation}
\bar{N} m_1 =  - \frac{1}{6} X_4 .
\end{equation}
We find, in agreement with \cite{Dymarsky:2011pm}, that the ADM mass is given by 
\begin{equation}\label{MTERM}
M = - \frac{48}{(2\pi)^4 \alpha'^4} \left[ \e^{4A+4p} A' - \frac{W}{3}\right] = \frac{8}{(2\pi)^4 \alpha'^4} X_4 + \mathcal{O}  \left(\bar{N}^2 \right).
\end{equation}
Implementing the UV expansions in \cite{Bena:2011wh} with $Q = 0$ leads to the same results with $\tau \rightarrow \infty$.
Hence, we see that the boundary term (\ref{BTERM}) and the ADM mass (\ref{MTERM}) coincide (up to a volume factor $v$).

\section{Conventions and useful equations}\label{AppConventions}

For the equations of motion of IIA/IIB SUGRA, we follow the conventions of \cite{Blaback:2010sj}. The duality relations between the field strengths are
\begin{equation}
\e^{\tfrac{5-n}{2}\phi} F_n=\star_{10} \ \sigma\left(F_{10-n}\right)\,,
\end{equation}
and we take
\begin{equation}
\delta_{9-p} = \sigma \left( \star_{9-p} 1 \right) \delta \left( \Sigma \right),
\end{equation}
where the operator $\sigma$ interchanges the indices or, equivalently, acts as
\begin{equation}
\sigma\left(A\right) = \left( -1 \right)^{\tfrac{n(n-1)}{2}} A\,,
\end{equation}
with $A$ an $n$-form.

We use the following conventions for the Hodge operator:
\begin{equation}
\left( \star_{10} A \right) \wedge A = |A|^2 \star_{10} 1 = \frac{1}{n!} A_{\mu_1 ... \mu_n} A^{\mu_1 ... \mu_n} \star_{10} 1\,.
\end{equation}
We also make repetitive use of
\begin{equation}
\star_{10} \left( A_{n} \wedge B_{m} \right) =\left(-1\right)^{n(9-p-m)} \star_{p+1} A_{n} \wedge \star_{9-p} B_{m}
\end{equation}
and
\begin{equation}
\star_{10} \star_{10} A_{n} = \left(-1\right)^{n(10-n)+1} A_{n} \,,
\end{equation}
where $A_{n}$ and $B_{m}$ are $n$- and $m$-forms in the $(p+1)$-dimensional worldvolume and $(9-p)$-dimensional transversal spaces, respectively.

The 1-forms used in the KS ansatz \eqref{ansatz-smearedks} satisfy
\begin{eqnarray}
&& \star_6 \left( g_1 \wedge g_2 \wedge g_5 \right) =  \e^{- 2y} \mathrm{d} \tau \wedge g_3 \wedge g_4 \,,\nonumber
\\
&& \star_6 \left( g_3 \wedge g_4 \wedge g_5 \right) = \e^{ 2y} \mathrm{d} \tau \wedge g_1 \wedge g_2 \,,\nonumber
\\
&& \star_6 \left[ \left( g_1 \wedge g_3 + g_2 \wedge g_4 \right) \wedge g_5 \right] = - \mathrm{d} \tau \wedge \left( g_1 \wedge g_3 + g_2 \wedge g_4 \right)\,
\end{eqnarray}
and
\begin{equation}
\mathrm{d} \tau \wedge \left( g_1 \wedge g_3 + g_2 \wedge g_4 \right) \wedge \left( g_1 \wedge g_3 + g_2 \wedge g_4 \right) \wedge g_5=- 2 \tilde \star_6\,.  
\end{equation}
Useful differentiation rules are
\begin{eqnarray}
&& \mathrm{d} \left( g_1 \wedge g_3 + g_2 \wedge g_4 \right) = \left( g_1 \wedge g_2 - g_3 \wedge g_4 \right) \wedge g_5 \,,\nonumber
\\
&& \mathrm{d} \left( g_1 \wedge g_2 \wedge g_5 \right) = \mathrm{d} \left( g_3 \wedge g_4 \wedge g_5 \right) = 0 \,,\nonumber
\\
&& \mathrm{d} \left( g_1 \wedge g_2 \right) = - \frac{1}{2} \left( g_1 \wedge g_3 + g_2 \wedge g_4 \right) \wedge g_5 \,,\nonumber
\\
&& \mathrm{d} \left( g_3 \wedge g_4 \right) = \frac{1}{2} \left( g_1 \wedge g_3 + g_2 \wedge g_4 \right) \wedge g_5\,.
\end{eqnarray}
Some useful formulas at zeroth order in $\bar{N}$ are
\begin{eqnarray}
&& \phi_0 = 0 \,,\nonumber
\\
&& \lambda_0 = \frac{1}{2 P} \left[ k'_0 \e^{2y_0} + f'_0 \e^{-2y_0} \right] = -1 \,,\nonumber
\\
&& \alpha_0 = - \e^{4 A_0 + 4 p_0 - 2 x_0 } = -\frac{1}{h(\tau)} \,,\nonumber
\\ 
&& \alpha'_0 = \e^{4 A_0 + 4 p_0 - 4 x_0 } \left[ f_0 \left( 2 P - F_0 \right) + k_0 F_0 \right] = - \e^{- 2 x_0 } \alpha_0 \left[ f_0 \left( 2 P - F_0 \right) + k_0 F_0 \right] \,,\nonumber
\\ 
&& \alpha'_0 =\alpha_0 \left( 4 A'_0 + 4 p'_0 - 2 x'_0\right) \,,\nonumber
\\
&& \frac{1}{2 P} \left[ k'_0 \e^{2y_0} - f'_0 \e^{-2y_0} \right] = 1 - \frac{F_0}{P} \,,\nonumber
\\
&& 0 = \e^{2x_0} A'_0 - \frac{1}{3} \e^{-6p_0}- \frac{1}{6}\e^{2 x_0 + y_0}- \frac{1}{6}\e^{2x_0 - y_0 }- \frac{1}{6}\left[ f_0 \left( 2 P - F_0 \right) + k_0 F_0 \right]\,.
\end{eqnarray}

\bibliography{anti}

\bibliographystyle{utphysmodb}

\end{document}